\documentclass[aps,superscriptaddress,amsmath,amssymb,floatfix,twocolumn,showpacs,amsfonts,longbibliography,reprint]{revtex4-2}

\usepackage{times}
\usepackage[varg]{txfonts}
\usepackage{textcomp}
\usepackage{graphicx}
\usepackage{subfigure}
\usepackage{color}
\usepackage[colorlinks=true,citecolor=blue,urlcolor=blue,linkcolor=blue,hyperindex]{hyperref}
\usepackage{braket}
\usepackage{float}

\usepackage{overpic}
\usepackage{bm}
\usepackage{amsbsy}
\usepackage{array}
\usepackage{booktabs}
\usepackage{tabularx}

\begin{document}

\title{Unraveling the Mn $L_3$-edge RIXS spectrum of lightly manganese doped Sr$_{3}$Ru$_{2}$O$_{7}$}
\author{Wei-Yang Chen}
\affiliation{State Key Laboratory of Optoelectronic Materials and Technologies, Guangdong Provincial Key Laboratory of Magnetoelectric Physics and Devices, Center for Neutron Science and Technology, School of Physics, Sun Yat-Sen University, Guangzhou 510275, China}
\author{Shih-Wen Huang}
\affiliation{Paul Scherrer Institut, Swiss Light Source, CH-5232 Villigen PSI, Switzerland}
\author{Yi Tseng}
\affiliation{Paul Scherrer Institut, Swiss Light Source, CH-5232 Villigen PSI, Switzerland}
\author{Wenliang Zhang}
\affiliation{Paul Scherrer Institut, Swiss Light Source, CH-5232 Villigen PSI, Switzerland}
\author{Eugenio Paris}
\affiliation{Paul Scherrer Institut, Swiss Light Source, CH-5232 Villigen PSI, Switzerland}
\author{Teguh Citra Asmara}
\affiliation{Paul Scherrer Institut, Swiss Light Source, CH-5232 Villigen PSI, Switzerland}
\author{Jenn-Min Lee}
\affiliation{MAX IV  Laboratory, Lund University, P. O. Box 118, 221 00 Lund, Sweden}
\author{Thorsten Schmitt}
\affiliation{Paul Scherrer Institut, Swiss Light Source, CH-5232 Villigen PSI, Switzerland}
\author{Yu-Cheng Shao}
\email[Corresponding author:]{shao.yc@nsrrc.org.tw}
\affiliation{National Synchrotron Radiation Research Center, Hsinchu 30076, Taiwan }
\affiliation{Advanced Light Source, Lawrence Berkeley National Laboratory, Berkeley, California 94720, USA}
\author{Yi-De Chuang}
\affiliation{Advanced Light Source, Lawrence Berkeley National Laboratory, Berkeley, California 94720, USA}
\author{Byron Freelon}
\affiliation{Department of Physics and Texas Center for Superconductivity, University of Houston, Houston, Texas 77204, United States}
\author{Dao-Xin Yao}
\email[Corresponding author:]{yaodaox@mail.sysu.edu.cn}
\affiliation{State Key Laboratory of Optoelectronic Materials and Technologies, Guangdong Provincial Key Laboratory of Magnetoelectric Physics and Devices, Center for Neutron Science and Technology, School of Physics, Sun Yat-Sen University, Guangzhou 510275, China}
\author{Trinanjan Datta}
\email[Corresponding author:]{tdatta@augusta.edu}
\affiliation{Department of Physics and Biophysics, Augusta University, 1120 15$^{th}$ Street, Augusta, Georgia 30912, USA}
\affiliation{Kavli Institute for Theoretical Physics, University of California, Santa Barbara, California 93106, USA}

\begin{abstract}
Resonant inelastic x-ray scattering (RIXS) experiment was performed at the Mn $L_3$ edge. A 10 $\%$ Mn-doped Sr$_{3}$Ru$_{2}$O$_{7}$ compound, where the Mn$^{3+}$ ions are in the 3$d^4$ state, were probed for $dd$ excitations. The dilute doping concentration allows one to treat the dopant Mn$^{3+}$ ions as effectively free in the host ruthenium compound. The local nature of $dd$ RIXS spectroscopy permits one to use a single-site model to simulate the experimental spectra. The simulated spectra reproduces the in-plane [100] experimental RIXS spectrum. We also predict the intensity for the in-plane [110] direction and the out-of-plane spin orientation configuration [001]. Based on our single-ion model we were able to fit the experimental data to obtain the crystal field parameters ($\Delta E_{e_{g}}$, $\Delta E_{t_{2g}}$), the $10Dq$ value, and the intra-orbital spin-flip energy 2$\mathcal{J}$(or $3J_{H}$, where $J_{H}$ is the Hund's energy) of the Mn$^{3+}$ ion. Utilizing our computed RIXS quantum transition amplitudes between the various $d$ orbitals of the Mn$^{3+}$ ion, the expression for the Kramers-Heisenberg cross section, and a self-consistent fitting procedure we also identify the energy boundaries of the non-spin-flip and spin-flip \textit{dd} excitations present in the experimental data. From our fitting procedure we obtain $2\mathcal{J} (3J_{H})=2.06$ eV, a value which is in excellent agreement with that computed from the free ion Racah parameters. We also identified the charge transfer boundary. In addition to predicting the microscopic parameters, we find a quantum spin-flip transition in the non-cross ($\pmb{\sigma}_{in}-\pmb{\sigma}_{out}$, $\pmb{\pi}_{in}-\pmb{\pi}_{out}$) x-ray polarization channels of the $dd$ RIXS spectra. A similar transition, was previously predicted ~\cite{xiong2020resonant} to occur in the $\pmb{\pi}-\pmb{\pi}$ channel of the magnon spectrum in the non-collinear non-coplanar Kagome compound composed of Cu$^{2+}$ 3d$^{9}$ ion.
\end{abstract}
\maketitle

\section{Introduction}\label{sec:introduction}
Quantum materials harbor electronic excitations which can be accurately fingerprinted using resonant inelastic x-ray scattering (RIXS) spectroscopy~\cite{RevModPhys.83.705}. The advantage of RIXS lies in its ability to probe a range of excitations that span low energy (phonons~\cite{PhysRevX.6.041019, PhysRevLett.110.265502, PhysRevLett.123.027001}, magnons~\cite{nmat3409, LeTacon2011, PhysRevLett.102.167401, Zhou2013, Lee2014, PhysRevB.103.L140409,xiong2020resonant}, bimagnons~\cite{Elnaggar2023, Schlappa2012} to high energy orbital $dd$ excitations~\cite{Moretti_Sala_2011, Huang2017, Bisogni2016}, charge transfer (CT) excitations~\cite{Bisogni2016, PhysRevX.6.041019, Hepting2018, PhysRevX.6.021020}, and plasmon~\cite{Gudarzi2023}). Often, RIXS provides a valuable complementary technique to angle-resolved photoemission (ARPES)~\cite{Zhang2022, Yang2018, Thompson2020}, Raman~\cite{GERLACH1930, PhysRevB.107.184402}, and inelastic neutron scattering (INS)~\cite{Tóth2016, Fong1999} spectroscopies. Transition metal \textit{L}-edge RIXS has been utilized to obtain insight into electronic orders~\cite{Wang2023, Calder2016, PhysRevB.101.024426}, electron-boson couplings~\cite{PhysRevResearch.2.023231, PhysRevResearch.4.033004}, (multi-)magnon excitations~\cite{xiong2020resonant, PhysRevB.85.214527}, and superconductivity~\cite{PhysRevLett.110.117005, Suzuki2018}. The enhanced cross-section under resonance condition gives RIXS the unique sensitivity to the chemical and bonding state of the probing element at or near the photon interaction site. This allows RIXS to be employed on materials that are not suitable for ARPES and INS studies, e.g. thin films or dilute dopants with small volume fractions, the buried interfaces~\cite{PhysRevB.83.201402}, and nano-particles~\cite{doi:10.1021/jp302847h} prepared under non-ultra-high vacuum conditions. Thus, RIXS is an excellent choice for exploring electronic behavior and mechanism, such as in high T$_{c}$ cuprates and colossal magnetoresistance (CMR) compounds like manganites. This experimental technique also finds popularity in other applied fields such as energy storage and novel catalysts~\cite{C7CP06786K}. 

In recent years, the condensed matter and materials science community have also gradually begun to use RIXS to probe the fundamental excitations of manganese compounds~\cite{PhysRevLett.94.047203, PhysRevB.78.155111, zhou2014glass, PhysRevLett.106.186404}. Since the discovery of CMR effect in manganese oxides ~\cite{chahara1993magnetoresistance, PhysRevLett.71.2331, 1995Giant, doi:10.1126/science.264.5157.413}, manganites have proven to be an important class of strongly correlated electronic materials, besides the high-temperature superconducting cuprates. In contrast to cuprates, CMR manganites (which have a perovskite structure) possess more interesting electronic composition and sophisticated electronic orders, such as CE-type charge, spin, and orbital order~\cite{maekawa2013physics}. 
\textit{K}-edge RIXS has been used to study electronic excitations in manganites exhibiting a range of ground states~\cite{PhysRevLett.94.047203}. Theoretical studies on RIXS has demonstrated its applicability to investigate the unique electronic excitations in orbital ordered manganites~\cite{PhysRevB.78.155111}. The antiferromagnetic and the CE-type spin ordering of Pr$_{x}$Ca$_{1-x}$MnO$_3$ has been studied  experimentally~\cite{zhou2014glass, PhysRevLett.106.186404}. 

In addition to manganese compounds, manganese ions as dilute dopant impurities can play an important role in influencing the magnetic properties of its host material. Consider the ruthenate Sr$_{3}$Ru$_{2}$O$_{7}$, where it has been experimentally found that Mn$^{3+}$ impurities can induce long-range magnetic ordering in the compound~\cite{PhysRevB.72.092404, PhysRevLett.101.016404, Panaccione_2011}. This is because the radial extension of the $4d$ wave function is significantly larger than that of the TM $3d$ or even O $2p$ orbitals. This leads to weaker correlation effect for the ruthenium oxide than for the $3d$ TM oxide. From a materials perspective, Sr$_{3}$Ru$_{2}$O$_{7}$ is considered as a metamagnetic metal on the verge of ferromagnetism. Because of the interaction between localized Mn $3d$ and delocalized Ru $4d$ - O $2p$ valence states, the Mn impurity exhibits a $3+$ oxidation state and a crystal field energy level inversion at room temperature~\cite{PhysRevLett.101.016404, Panaccione_2011}. At low temperatures, a metal-insulator transition can be measured, whose transition temperature rises with increasing Mn$^{3+}$ doping~\cite{PhysRevB.72.092404}. When the Mn$^{3+}$ doping reaches 10$\%$ one can detect a $(1/4,1/4,0)$ antiferromagnetic order in the Mn-doped Sr$_3$Ru$_2$O$_7$~\cite{PhysRevB.86.041102, PhysRevResearch.3.043183, Leshen2019}. In this article, we utilize Mn $L_3$-edge RIXS to investigate the local nature of the electronic excitations of the Mn$^{3+}$ ions in strontium ruthenate.

In the lightly manganese doped ruthenate, the Mn$^{3+}$ ions can be viewed as free ions in a spin-ordered environment. A similar single-ion model of Cu$^{2+}$ was used to probe the $dd$ excitations in cuprates~\cite{Moretti_Sala_2011}. Since we are not probing the Ru \textit{L} edge, the effect of Ru is not visible in the spectra. Therefore, we can unravel the $dd$ excitations of the single Mn$^{3+}$ ion by RIXS. Thus, we can claim that we are measuring the Mn impurity system where Mn is the minority dopant in otherwise the Ru majority network. So these Mn dopants should be treated as isolated atoms. In that case, the non-interacting picture (e.g. not considering nearest Mn neighbors and their influence via Mn-O-Mn hybridization) should be a viable perspective. This is our single-ion model. The local nature of RIXS spectroscopy often permits the use of single-site approximation  to simulate the experimental spectrum. The simulated spectrum are mostly compared with experimental ones where the probed elements are the dominant species. It is still interesting to examine how far one can extend this kind of local treatment, and the dilute dopant systems are the ideal candidates for such studies.

In this paper, using a single-ion model, we theoretically calculate the RIXS cross sections for all possible \textit{dd} excitations of the Mn$^{3+}$ ions under both the cross and the non-cross x-ray polarization channels. From the comparison of the RIXS simulation spectra with the experimental data, we obtain the crystal field parameters and intra-orbital spin-flip energy of Mn$^{3+}$ ions in the host material. We also estimate the energy boundaries of the non spin-flip (NSF),  spin-flip (SF), and charge-transfer excitations. We predict the RIXS spectra for spin orientation that changes from the in-plane to the out-of-plane configuration. As the spin orientation changes from in-plane to out-of-plane, we find non-trivial intra-orbital SF excitation channels for both non-cross polarizations ($\pmb{\sigma}$, $\pmb{\pi}$).

This article is organized as follows. In Sec.~\ref{sec:experiment}, we show the experimental methods, scattering geometry and experimental RIXS spectra. In Sec.~\ref{sec:model}, we introduce the single-ion model of Mn$^{3+}$  in cubic perovskite structure with $D_{4h}$ symmetry. In Sec.~\ref{sec:calculation}, we calculate the transition amplitudes of all possible Mn$^{3+}$ $dd$ excitations under different polarization channels. In Sec.~\ref{sec:parameter}, we obtain energy parameters by the characteristic peaks of experimental spectra. In Sec.~\ref{sec:rixs spectrum}, we use transition amplitudes and energy parameters to simulate experimental data and predict RIXS spectra for other experimental conditions. In Sec.~\ref{sec:conclusion}, we provide our conclusions. 

\begin{figure}[t]
\centerline{\includegraphics[width=9.0cm]{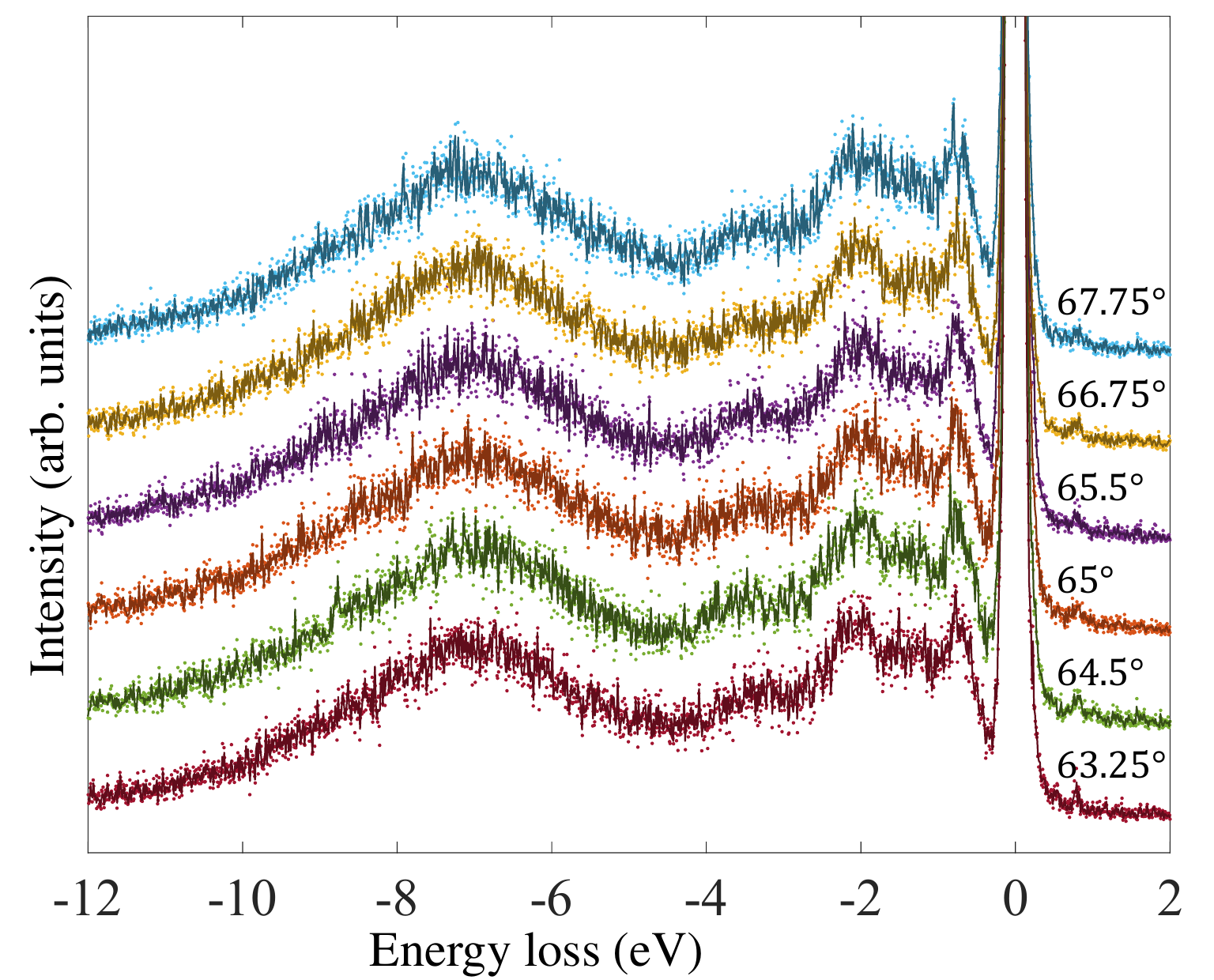}}
\caption{Original (dot) and 2X binned (solid) experimental RIXS spectrum of a 10\% Mn-doped Sr$_3$Ru$_2$O$_7$ recorded at Mn $L_{3}$ resonance ($\approx$ 641.5 eV) with a linear vertical photon polarization ($\pmb{\sigma}$ polarization). The scattering angle (2$\theta$) is 130$^{\circ}$ and the sample angle ($\theta$) with respect to its surface is listed in the figure. The measurement temperature is 25 K, below the spin ordering transition temperature. The features up to 4 eV energy loss with respect to the elastic peak (0 eV energy loss) are the $dd$ excitations in Mn$^{3+} d^{4}$ structure, and the broad hump around 7 eV energy loss is the charge-transfer excitation.}
\label{fig: experiment from -12 to 2}
\end{figure}

\begin{figure*}[t]
\centerline{\includegraphics[width=18cm]{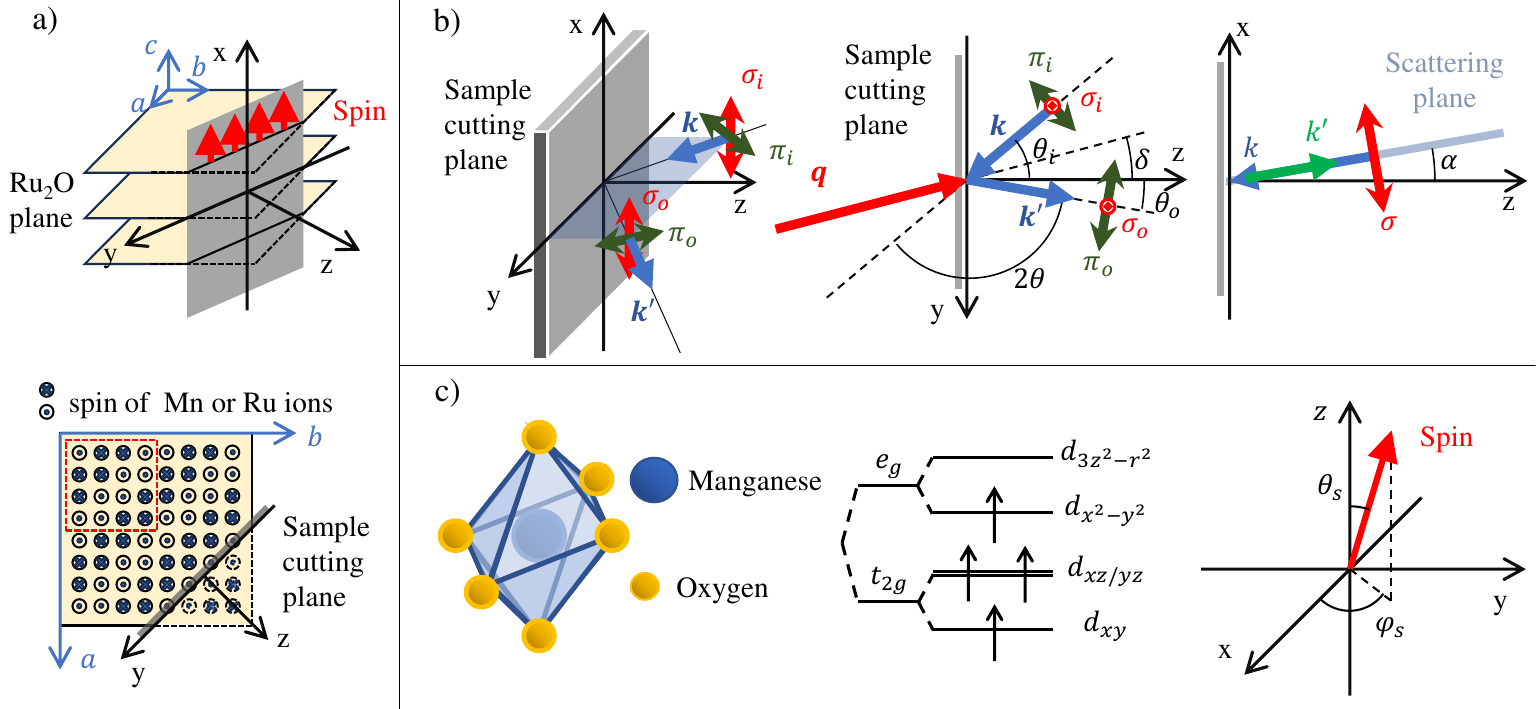}}
\caption{Crystal plane, X-ray scattering geometry, polarization configuration of the beam,local cluster, $d$-orbital energy level, and spin orientation of the sample in the experimental set-up.  a) The crystal axes are denoted by $(a,b,c)$, where Ru$_{2}$O plane stacking along the c-axis. The crystal plane of the experimental sample is cleaved in the (110)$_{\text{abc}}$ direction indicated by the shaded plane, which is used as scattering surface. The local spin orientations should all lie along the c-axis, up or down depending on which diagonal row is exposed. The cutting plane of the cleaved sample establishes the scattering geometric axes denoted by (x, y, z), where the xy plane is parallel to the sample surface and the z axis is perpendicular to it. b) The scattering geometry axes are denoted by $(x,y,z)$, where the $xy$ plane is parallel to the sample surface and the $z$ axis is perpendicular to it. The incoming (outgoing) wavevector $\bm{k}~(\bm{k}^{\prime})$ and the corresponding incoming (outgoing) scattering angle relative to the z-axis is given by $\theta_i~(\theta_o)$. The scattering angle between $\bm{k}$ and $\bm{k}^{\prime}$ is $2\theta=130^{\circ}$ (for all our cases). The scattering wavevector $\bm{q}=\bm{k}^{\prime} -\bm{k}$ makes an angle $\delta$ relative to the z-axis. The in-plane (out-plane) x-ray polarization is given by $\pmb{\pi}~(\pmb{\sigma})$. Additionally, as a result of imperfect experimental manipulation, the cutting plane is not actually perfectly perpendicular to the scattering plane. Thus, the tilted angle $\alpha$ between z-axis and scattering plane is about 6$^{\circ}$ to 9.5$^{\circ}$. c) Octahedral cluster of a central manganese and six surrounding ligand oxygen with a stretched $c$-axis and the effect of the octahedral crystal field on the energy levels of the Mn$^{3+}$ $3d$ electrons in a cubic perovskite structure with D$_{4h}$ symmetry. The general spin orientation angles relative to the sample axis are given by $(\theta_s,\phi_s)$. Therefore, in our experimental case $\theta_{s}=\pi/2$ and $\phi_{s}=0$ in scattering process.}
\label{fig: scattering geometry}
\end{figure*}

\section{Experimental methods and RIXS spectrum}\label{sec:experiment}

Single crystal Sr$_{3}$(Ru$_{1-x}$Mn$_{x}$)$_{2}$O$_{7}$ was grown using the traveling solvent float zone method \cite{MathieuPhysRevB.72.092404}. It was cut along the (110) direction (I4/mmm notation) and polished with 100 nm diamond paste to expose a mirror-like surface for the experiments. Similar preparation method was used in the previous publication \cite{SumanHossain2013}. Prior to the high-resolution RIXS experiment, the sample was measured at the qRIXS endstation at beamline 8, Advanced Light Source (ALS), Lawrence Berkeley National Laboratory (LBNL), to confirm the presence of electronic order similar to the previously reported results \cite{SumanHossain2013}. The RIXS experiments were conducted at ADRESS beamline, Swiss Light Source (SLS), Paul Scherrer Institut (PSI)~\cite{Strocov2010-ha}. The sample was mounted on a cryogenic manipulator with temperature control from 25 K to room temperature. Most measurements were conducted below 100 K, acrossing the electronic ordering temperature of $\sim 60$ K in this sample. A linear vertical polarization ($\pmb{\sigma}$) was used to enhance the elastic signal for the spin order. In Fig.~\ref{fig: experiment from -12 to 2}, we show the Mn $L_{3}$-edge RIXS spectrum of a 10\% Mn doped Sr$_{3}$Ru$_{2}$O$_{7}$ sample as an example. Although the spectrum exhibits a higher noise level, the inter-orbital ($dd$) excitations up to $\approx  4$ eV energy loss remain visible and displays a distinct spectrum line shape compared to La$_{0.5}$Ca$_{0.5}$MnO$_{3}$~\cite{doi:10.1142/S0218625X02003196}. Here onwards, for visual clarity we display the 2X binned experimental data plots in rest of the article. Note, the fitting was performed on the original data (see Sec.~\ref{sec:parameter} for a detailed description of the method). The less-well defined lineshape indicates an intrinsic broadening likely caused by a reduced final state lifetime for these excitations. 

In Fig.~\ref{fig: scattering geometry} we show the crystal plane and the spin orientation of the sample in the experimental setting, the x-ray scattering geometry, and the polarization configuration of the incoming and outgoing x-ray beam that has been used both in the experiment and in our calculation. In Fig.~\ref{fig: scattering geometry}(a), the Ru$_{2}$O plane cleaved in the (110)$_{\text{abc}}$ direction is used as scattering surface. The determination of the coordinate system depends on the cutting plane of the sample which is aligned along the diagonal of the supercell. Since the layered samples are stretched along the x-axis, the spin orientation ($\theta_{s}=\pi/2,~\phi_{s}=0$) is pointed along that direction as well. The sample orientation with respect to the incident photon Poynting vector is shown in Fig.~\ref{fig: scattering geometry}(b). In the ideal situation, when the tilt angle $\alpha$ between the scattering plane and the yz plane is zero, the incident (scattered) polarization $\pmb{\epsilon}_{in}$ ($\pmb{\epsilon}_{out}$) either can be $\pmb{\sigma}$ or $\pmb{\pi}$. For example, $\pmb{\sigma}_{in}=\pmb{\sigma}_{out}=(1,0,0),~\pmb{\pi}_{in}=(0, \cos \theta_{i}, \sin\theta_{i})$, and $\pmb{\pi}_{out}=(0, \cos\theta_{o}, -\sin\theta_{o})$. The incident (scattered) angle $\theta_{i}$($\theta_{o}$) is determined by the scattering plane for the experimentally controlled variable angle $\delta$ and the fixed scattering angle $2\theta =130^{\circ}$. This can be attributed to the fact that imperfections in experimental cutting of the sample results in an angle $\alpha$ of about 6$^{\circ}$ to 9.5$^{\circ}$ between the two planes. Thus, to appropriately model the experimental scattering geometry as shown in Fig.~\ref{fig: scattering geometry}(b), we employ rotation matrix $R(\alpha)$ around the $b$-axis that is given by \begin{equation}
    R(\alpha )=\left[
    \begin{array}{ccc}
    \cos\alpha &0  & \sin\alpha \\
    0& 1 & 0 \\
    -\sin\alpha & 0 & \cos\alpha 
\end{array}\right]
\label{eq: rotation}
\end{equation} In the experiment the incoming polarization was set to $\pmb{\sigma}$ and the outgoing polarization was summed over both the polarization directions. Hence, the polarization of the incident and scattered x-ray beams can be expressed as $\pmb{\epsilon}_{in} =R(\alpha)\pmb{\sigma}_{in}$, $\pmb{\epsilon}_{out} =R(\alpha)\pmb{\sigma}_{out}$, or $\pmb{\epsilon}_{out} =R(\alpha)\pmb{\pi}_{out}$. 

\begin{figure*}[t]
\centerline{\includegraphics[width=18.0cm]{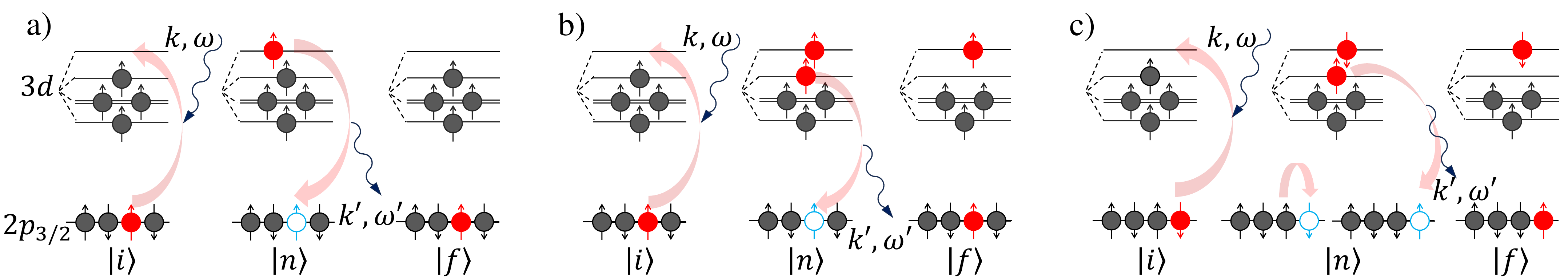}}
\caption{The schematic diagram of various $dd$ excitation scattering processes between Mn-ion and x-ray photon. a) In the elastic scattering channel the final state is equal to the initial state (non spin-flipped). b) Non spin-flip inelastic scattering process. c) Spin-flip inelastic scattering process. In the intermediate state the spin of core hole flips because of the spin-orbit coupling present in the 2p orbital. Therefore, SF $dd$ excitation process is realized. }
\label{fig: scattering type}
\end{figure*}

\section{Single-ion model}\label{sec:model} The $3d^{4}$ electronic configuration of manganese ions in manganites belong to the Mn$^{3+}$ oxidation state. The $c$-axis stretched coordination complex which is formed by the Mn$^{3+}$ ions and the six surrounding oxygen ligands at the vertices of an octahedron form a cubic perovskite structure. Therefore, the degeneracy of the $e_g$ orbitals (composed of $d_{x^2-y^2}$ and $d_{3z^2-r^2}$) and the $t_{2g}$ orbitals (composed of $d_{xy}$, $d_{yz}$, and $d_{zx}$) splits under the influence of the $D_{4h}$ crystal field (CF) symmetry, see Fig.~\ref{fig: scattering geometry}(b). We implement a single-ion model to simulate the resonant inelastic x-ray scattering (RIXS) $dd$ excitation spectrum experimental data of the Mn-doped Sr$_{3}$Ru$_{2}$O$_{7}$ compound at the Mn $L_3$ edge resonance (described in previous section, also see Fig.~\ref{fig: experiment from -12 to 2}). On basis of this model, we can compute the local $dd$-transitions between the different $d$ orbitals, that is, the non-degenerate $3d$ electron levels of the same ion, with energies which are of the order of electron volt (eV) for transition metal oxides. The consideration of the single-ion approximation is adequate due to the local nature of the $dd$ excitation. 

We consider the octahedral cluster of a Mn$^{3+}$ ion and the six surrounding oxygen ligands. The $|d^{\tau}_{\alpha}\rangle$ orbital wave function of the Mn-ion is obtained from a linear combination of the $|Y^{~2}_{\pm m}\rangle$ spherical harmonics where $\alpha$ represents a $d$-orbital index $(3z^2 - r^2, x^2 - y^2, xz, yz, xy)$ with spin $\tau$ ($\pm~1/2$). The $2p$ ligand states are spin–orbit coupled where $\mathbf{L}\cdot \mathbf{S}=\frac{1}{2}(L^{+}S^{-}+L^-S^+)+L^zS^z$. The wave function of the $2p_{3/2}$ electrons at the $L_{3}$-edge can be written as a linear combination of the spin-up and spin-down states given by $\left|p_{3/2,m}\right\rangle=C^{3/2,m}_{m_l,m_s}\left|Y^1_{m_l}\right\rangle\otimes\left|m_s\right\rangle$, where $C^{3/2,m}_{m_l,m_s}$ are the Clebsch-Gordan coefficients associated with magnetic ($m=m_l+m_s$), orbital ($m_l=\pm 1$, and $0$) and spin ($m_s= \pm 1/2$) quantum numbers. The calculations in this paper can distinguish between the SF and NSF $dd$ excitation channels.

\section{RIXS transition amplitudes}\label{sec:calculation}

In general, RIXS is visualized as an inelastic scattering of x-ray photons with clusters of matter ions. First, the incident x-ray photon resonantly excites a core-electron into an unoccupied state, (for e.g., an energy level or a band). This intermediate state is in a non-equilibrium situation with a core-hole and an excited electronic state with ultra-short survival times of the order of femtoseconds. Afterwards, the interplay of the core-hole and the excited electron, accompanied by the simultaneous emission of the scattered photons, relaxes the system either to its initial state (elastic scattering) in Fig.~\ref{fig: scattering type}(a), or to a different low-energy excited state (inelastic scattering) in Figs.~\ref{fig: scattering type}(b) and (c). The elastic scattering process is a NSF channel. But, the inelastic scattering process can be accompanied either with a SF or a NSF channel. Such a direct RIXS process can be considered as a second-order process which is described by the Kramers-Heisenberg equation~\cite{blume1985magnetic, schulke2007electron, kramers1925streuung, sakurai1967advanced}. At zero temperature the double-differential RIXS scattering cross section is given by \begin{equation} \frac{d^{2}\sigma }{d\omega d\Omega }\propto\sum_{f}^{} \left|\sum_{n}^{}\frac{\left\langle f \right |\pmb{\epsilon}_{out}\cdot\mathbf{D}^{\prime\dagger}\left|n\right\rangle\left\langle n\right|\pmb{\epsilon}_{in} \cdot\mathbf{D}\left|i\right \rangle}{\Delta E +\hbar\omega_{k} +i\Gamma} \right|^{2}\delta\left(\Delta E+ \hbar\omega \right),\label{eq: Rixs cross}\end{equation} where $\Delta E = E_{i}-E_{n}$ is the energy difference between the initial state $\left|i\right\rangle $ and the final state $\left|f\right\rangle $, $\left|n\right\rangle $ is the intermediate state with energy $E_{n}$, and $\hbar\omega$ is the transferred photon energy. The incident (scattered) polarization vector dependence is given by $\pmb{\epsilon}_{in}$ ($\pmb{\epsilon}_{out}$). The transition operator $\mathbf{D}$ in the dipole approximation is given by $\mathbf{D}=\frac{1}{\sqrt{N}}\sum_{i=1}^{N}e^{i\mathbf{k} \cdot \mathbf{R} _i} \mathbf{D}_i$, where $\mathbf{D}_i=e^{i\mathbf{k} \cdot \mathbf{r} _i}\mathbf{p}_i\approx\mathbf{p}_i\approx\mathbf{r}_i$. In this equation we expressed the $\mathbf{D}$ operator in terms of the local transition operator $\mathbf{D}_{i}$, which is further simplified as $\mathbf{r}_{i}$.
 
\begin{figure*}[b]
\centerline{\includegraphics[width=18cm]{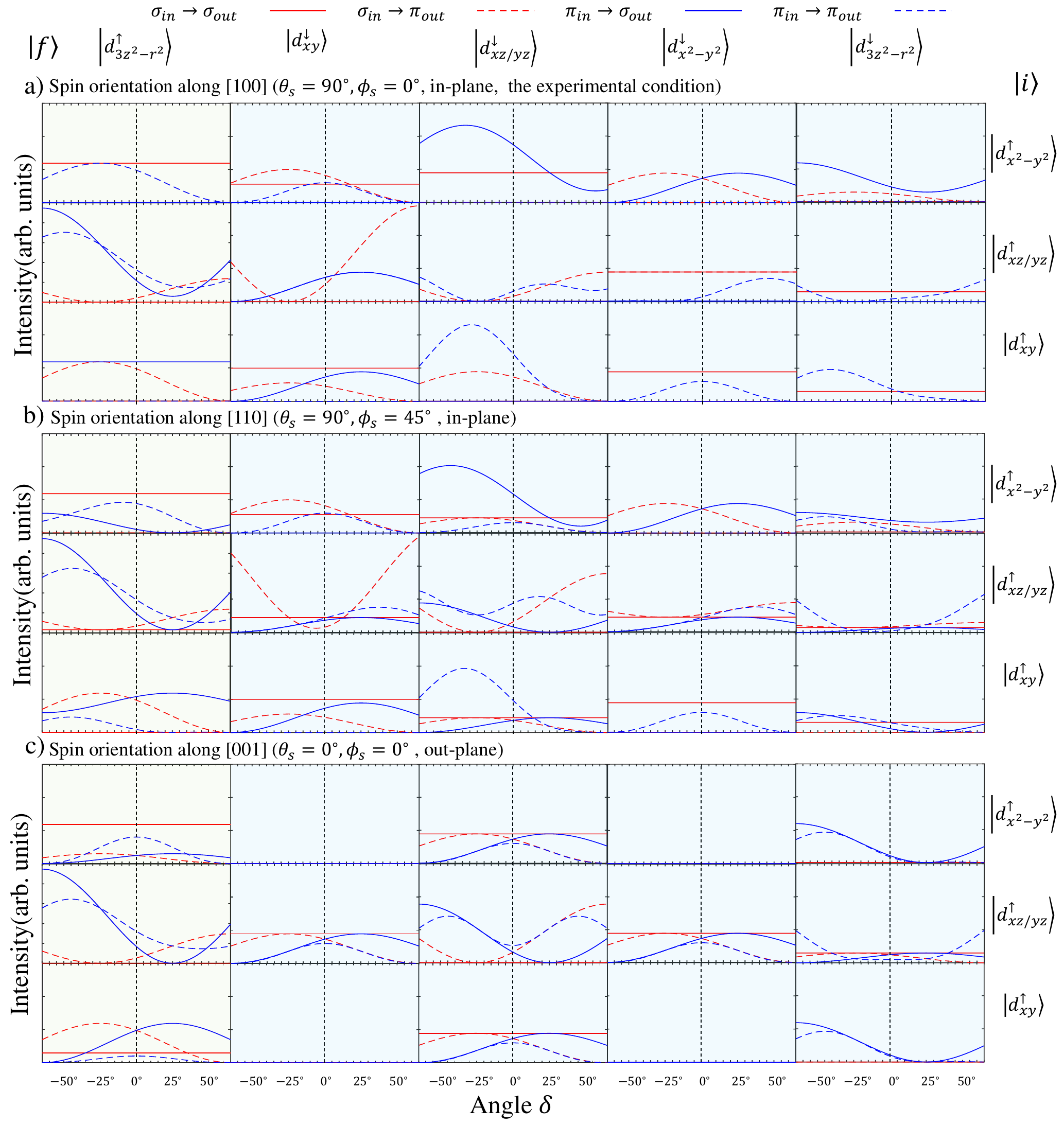}}
\caption{The transition amplitudes of Mn$^{3+}$ ion for single polarization  channels of all possible $dd$ excitation when the scattering angle (2$\theta$) is 130$^{\circ}$. Noteworthy, in the manganese ion, there are four electrons corresponding to three horizontal columns of electronic ground states, and six holes corresponding to five vertical columns of low-energy excited states. }
\label{fig:amplitudes}
\end{figure*} 

In Fig.~\ref{fig: scattering type}, we outline the microscopic quantum processes that contribute to the RIXS intensity of the experimental spectra. First, a $2p_{3/2}$ electron is resonantly promoted into the $3d$ state by absorption of a photon. The second step is the radiative excitation of a $3d$ electron into the $2p_{3/2}$ levels. This process can be written as the following sequence of events $-$ 2p$^{4}_{3/2}$3d$^{4}$$\rightarrow$2p$^{3}_{3/2}$3d$^{5}$$\rightarrow$2p$^{4}_{3/2}$3d$^{4*}$, where the * indicates the final state. Depending on the final state, we can also divide this scattering process into three types, see Fig.~\ref{fig: scattering type}. In the case of inelastic scattering process, the Mn$^{3+}$ ion is left behind in a low-energy excited state which may undergo a SF or not. In Fig.~\ref{fig: scattering type}(b) we show a NSF $dd$ excitation channel. However, the effects of spin-orbit coupling on the core hole of the $2p$ state can create a SF in the intermediate of the third process, see Fig.~\ref{fig: scattering type}(c). Using these quantum transitions as the basic ingredients, we simulated the Mn$^{3+}$ $L_{3}$-edge experimental RIXS intensity. Since, we are interested in modeling the $dd$ regime, we ignored the elastic contribution, and considered only the inelastic NSF and the SF $dd$ scattering channels. The transition amplitudes of the $dd$ excitation RIXS spectrum from $d^{\tau^{\prime}}_{\alpha}$ to $d^{\tau}_{\beta}$ can be written as \begin{equation}W^{\pmb{\epsilon}_{out}}_{\pmb{\epsilon}_{in}}\left(d^{\tau^{\prime}}_{\alpha},d^{\tau}_{\beta}\right)\propto\left|\sum_{m}^{}\left \langle d^{\tau}_{\beta}\right|\pmb{\epsilon}_{in}\cdot\mathbf{r}\left|2p_{3/2,m}\right\rangle\left\langle 2p_{3/2,m}\right|\pmb{\epsilon}_{out}\cdot\mathbf{r}^{\prime\dagger}\left|d^{\tau^{\prime}}_{\alpha}\right\rangle\right|^{2}, \label{eq: amplitude}
\end{equation} where $m=\pm 3/2$ and $\pm 1/2$. 

In Fig.~\ref{fig:amplitudes} we show the transition amplitudes from three initial states $|d^{\uparrow}_{x^2-y^2}\rangle,|d^{\uparrow}_{xz/yz}\rangle$, and $|d^{\uparrow}_{xy}\rangle$ to five final states $(|d^{\uparrow}_{3z^2 - r^2}\rangle, |d^{\downarrow}_{xy}\rangle, |d^{\downarrow}_{xz/yz}\rangle, |d^{\downarrow}_{x^2-y^2}\rangle)$, and $|d^{\downarrow}_{3z^2 - r^2}\rangle$ corresponding to inelastic NSF (first column) and inelastic SF (remaining four columns) $dd$ excitation channels. We compute the RIXS transition amplitudes for the sample spin orientation utilized in the experimental set-up (along [100]) and for a couple more theoretical cases (along [001] and [110]) crystal axes. Our calculation considers all possible SF and NSF $dd$ excitation channels in the experimental spectrum near 0 eV, see Fig.~\ref{fig: experiment from -12 to 2}. We also compute the RIXS transition amplitudes for various polarization channels. This includes the non-cross $(\pmb{\sigma}_{in}-\pmb{\sigma}_{out}, \pmb{\pi}_{in}-\pmb{\pi}_{out})$ channels and the cross polarization channels $(\pmb{\sigma}_{in}-\pmb{\pi}_{out}, \pmb{\pi}_{in}-\pmb{\sigma}_{out})$. The $dd$ excitation channels of the single-ion Cu$^{2+}$ $L_3$ edge RIXS cross section has been investigated~\cite{Moretti_Sala_2011, RevModPhys.83.705}. The study analyzed one elastic channel and seven inelastic channels. Compared to the $3d^9$ state of the (undoped) cuprates, the Mn$^{3+}$-ion exhibits fifteen inelastic $dd$ excitation channels due to the complexity of the $d^{4}$ electronic energy levels. $dd$ excitation channels which are mutually shared by both the Cu and the Mn ion, share similar spectrum. But, the presence of additional electronic transition pathways in Mn causes it to display behaviour which is unique to its situation. 

In Fig.~\ref{fig:amplitudes}, focusing on $\pmb{\sigma}$ as the incoming x-ray beam orientation (guided by the experiment), we find from Figs.~\ref{fig:amplitudes}(a) - \ref{fig:amplitudes}(b) (in-plane spin orientation) that the intra-orbital SF contribution to the RIXS intensity can arise from one non-cross $\pmb{\sigma}_{in}-\pmb{\sigma}_{out}$ channel and three $\pmb{\sigma}_{in}-\pmb{\pi}_{out}$ cross channel. While the cross polarization SF channel non-zero transition amplitude agrees with the common wisdom of angular momentum conservation, the intra-orbital SF transition occuring in the non-cross $\pmb{\sigma}_{in}-\pmb{\sigma}_{out}$ channel for the in-plane situation (as in the experiment) is non-trivial. When the polarization of the outgoing and incoming photons is the same the change in angular momentum of the photons is zero ($\Delta \mathbf{L}_b = 0$). If the orbitals of the transition channel remain unchanged ($\Delta \mathbf{L} = 0$), spin flipping should not occur ($\Delta \mathbf{S} = 0$). However, here we find both experimentally and theoretically, the opposite happens. In Fig.~\ref{fig:amplitudes}(c) there are four SF transition possibilities which occur with zero amplitudes ($d_{xy}$ and $d_{x^2-y^2}$ and vice-versa). Additionally, we observe that the only one intra-orbital SF amplitude, consistent with conservation of angular momentum, arises from the $\pmb{\sigma}_{in}-\pmb{\pi}_{out}$ cross polarization channel. 

The above (apparent) violation of conservation of angular momentum in the RIXS transition amplitudes has been predicted earlier by a couple of the authors in this paper~\cite{xiong2020resonant} and and observed experimentally \cite{PhysRevB.99.134517} in Cu$^{2+}$ magnon excitation. Interestingly, we find a similar behavior in the intra-orbital SF excitation of Mn$^{3+}$-ion. Note, even though both the magnon and the $dd$ excitation are composed of the intra-orbital spin-flip operator, in the case of magnons, this intra-orbital spin-flip operator is superposed with the wave vector and gives rise to the magnon dispersion. $dd$ excitation can be localized (as in our treatment) or it can be delocalized~\cite{Schlappa2012, PhysRevLett.103.107205} which can give rise to possible orbitons in orbital ordered materials. In both the localized $dd$ and the magnon excitation cases we find non-trivial non-zero SF amplitudes in the non-cross x-ray polarization channels. But, there are some subtle differences. In the case of Cu$^{2+}$ ion in the Kagome compound with $D_{4h}$ symmetry, the spins were in a non-collinear and non-coplanar arrangement (essentially one can think of this as out-of-plane in the language of this article). There the $\pmb{\pi}_{in}-\pmb{\pi}_{out}$ channel had the non-trivial non-zero intensity. However, in our case for the Mn$^{3+}$, the non-trivial intra-orbital SF happens in the in-plane spin orientation for the $\pmb{\sigma}_{in}-\pmb{\sigma}_{out}$ channel. But, when the spin is out-of-plane, the non-trivial intra-orbital SF happens in the $\pmb{\pi}_{in}-\pmb{\pi}_{out}$ channel.

To explain the above apparent transitions we will track the angular momentum changes in the x-ray beam $(\Delta \mathbf{L}_b)$, orbital angular momentum $\Delta \mathbf{L}$, and the spin angular momentum $\Delta \mathbf{S}$. Considering the initial and the final states of the $3d$ orbital during one such apparent non-cross polarization channel transition, the total angular momentum $\Delta \mathbf{L}_b + \Delta \mathbf{L} + \Delta \mathbf{S}$ ($\ne 0$ because of the spin-flip ) $\ne 0$, does not appear to be conserved. However, when the changes of the intermediate states and the angular momentum of the $2p$ orbitals are considered carefully, we can rationalize the scattering occurring in these channels. The variation of the angular momentum can be rewritten as $\Delta \mathbf{L} = \Delta \mathbf{L}_{3d} + \Delta \mathbf{L}_{2p}$ and $\Delta \mathbf{S} = \Delta \mathbf{S}_{2p}$, because the variations of the total spin angular momentum is caused by the spin-flip on the $2p$ orbital. In Fig.~\ref{fig: scattering type}(c), the spin of core hole in $2p$ orbital flips ($\Delta \mathbf{S}_{2p} \ne 0$) because of the spin-orbital coupling ($\mathbf{L}\cdot \mathbf{S}= 1/2(L^{+}S^{-}+L^-S^+)+L^z S^z$) in the intermediate state. In this process, the angular momentum of $2p$ orbital remains constant ($\Delta \mathbf{J}_{2p} = \Delta \mathbf{L}_{2p} + \Delta \mathbf{S}_{2p} =0 $). Therefore, these intra-orbital non-trivial SF transition in the non-cross ($\pmb{\sigma}_{in}-\pmb{\sigma}_{out}$, $\pmb{\sigma}_{in}-\pmb{\sigma}_{out}$) x-ray polarization channels can satisfy the angular momentum conservation ($\Delta \mathbf{L}_b + \Delta \mathbf{L}_{3d} + \Delta \mathbf{L}_{2p} + \Delta \mathbf{S}_{2p} = 0 $).

\section{Crystal field and energy level fitting procedure}\label{sec:parameter}

\begin{figure*}[t]
\centerline{\includegraphics[width=18.0cm]{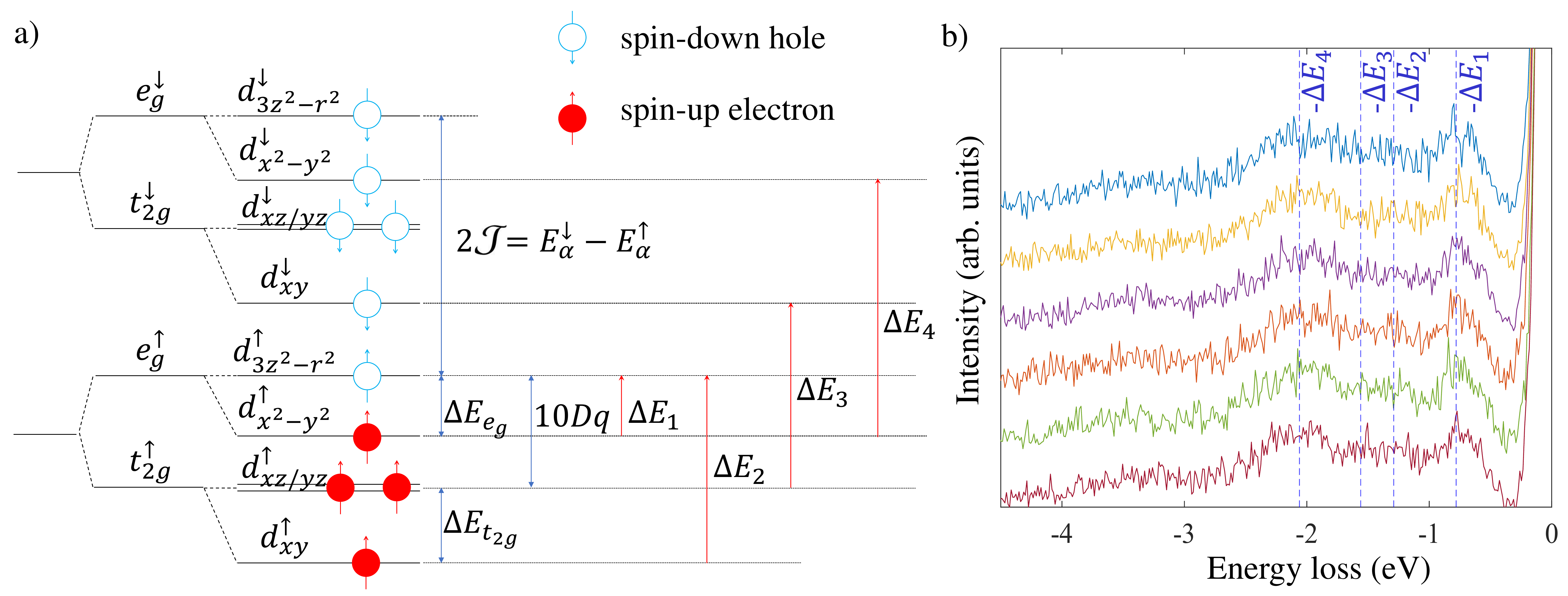}}
\caption{Crystal field parameter definitions. b) Identification of the characteristic peak energy in the experimental RIXS spectrum.}
\label{fig:crystal feild}
\end{figure*}

In this section we utilize the $dd$ formulation described above to extract the energy levels of the Mn$^{3+}$ ion in the host compound, the crystal field parameters, and the Hund's energy. These quantities play an important role in deciding the final appearance of the RIXS scattering cross section, see Eq.~\eqref{eq: Rixs cross}. Based upon the parameters extracted from the experimental data, we calculate the direct RIXS cross section using the local transition amplitudes. The energies of the initial and final states are directly associated with the orbital state occupied by the electrons. According to Fig.~\ref{fig:crystal feild}(a), the Mn$^{3+}$ ions in the $d^{4}$ electronic configuration possess four electron-occupied states and six hole-occupied states. To simulate the RIXS spectra we perform the following steps.

In the first step, we identify the characteristic peaks $\Delta E_i~(i = 1, \dots, 4)$ occurring in the experimental data shown in Fig.~\ref{fig:crystal feild}(b). The range over which these peaks vary are shown in Table~\ref{table:energy level}. Next, we use the transition amplitude results shown in Fig.~\ref{fig:amplitudes} and the definition of the Mn$^{3+}$ ion energy levels displayed in Fig.~\ref{fig:crystal feild} to determine the excitation channels corresponding to the characteristic peaks. The very first inelastic peak which has the smallest energy loss is $\Delta E_1$. Therefore, this peak corresponds to the excitation channel with the smallest energy difference between the initial and final states, where the electron is excited from $|d^{\uparrow}_{x^2-y^2}\rangle$ to $|d^{\uparrow}_{3z^2-r^2}\rangle$. The second possible transition is from $|d^{\uparrow}_{xz/yz}\rangle$ to $|d^{\uparrow}_{3z^2-r^2}\rangle$. However, under the experimental conditions ($\delta \approx 0$ and $\pmb{\epsilon}_{in}=R(\alpha)\pmb{\sigma}_{in}$), this particular excitation channel occurs with relatively small transition amplitude. Thus it does not have a significant contribution to the characteristic peaks of the total RIXS spectrum. We ignore this excitation channel in our calculation. The second peak $\Delta E_2$, that we have identified, represents the excitation from $|d^{\uparrow}_{xy}\rangle$ to $|d^{\uparrow}_{3z^2-r^2}\rangle$. The third peak $\Delta E_3$ is the SF $dd$ excitation channel from $|d^{\uparrow}_{xz/yz}\rangle$ to $|d^{\downarrow}_{xy}\rangle$. The fourth peak $\Delta E_4$ generates the largest RIXS intensity signal. This is caused by multiple SF $dd$ excitations between $d$-orbitals of the same type, that is $|d^{\uparrow}_{\alpha}\rangle \rightarrow |d^{\downarrow}_{\alpha}\rangle $. Thus, the energy losses corresponding to these four characteristic peaks can be related to the energy levels and crystal field parameters as \begin{subequations}
\begin{align}
&\Delta E_{1} = E^{\uparrow}_{3z^2-r^2} - E^{\uparrow}_{x^2-y^2} = \Delta E_{e_g}, \label{eq:e1}\\
&\Delta E_{2} = E^{\uparrow}_{xy} - E^{\uparrow}_{3z^2-r^2} = \Delta E_{e_g} + 10Dq, \label{eq:e2}\\
&\Delta E_{3} = E^{\downarrow}_{xy} - E^{\uparrow}_{xz/yz} = 2\mathcal{J} - \Delta E_{t_{2g}}, \label{eq:e3}\\
&\Delta E_{4} = E^{\downarrow}_{\alpha} - E^{\uparrow}_{\alpha} = 2\mathcal{J}. \label{eq:e4}
\end{align}
\end{subequations} This procedure allows us to greatly narrow the range of parameter choices and assists in reducing computational overhead associated with the fitting procedure. 
\begin{table}[b]
\begin{tabular}{ p{2cm}<{\centering} p{2cm}<{\centering} p{2cm}<{\centering} p{2cm}<{\centering} }
 \hline
 \multicolumn{4}{c}{Characteristic peaks of experimental RIXS spectrum (eV)} \\
 \hline
 $\Delta E_{1}$ & $\Delta E_{2}$ & $\Delta E_{3}$ & $\Delta E_{4}$\\
 \hline
0.73$\sim$0.83 & 1.24$\sim$1.34 & 1.51$\sim$1.61 & 1.96$\sim$2.16 \\ 
 \hline
\end{tabular}
\begin{tabular}{ p{2cm}<{\centering} p{2cm}<{\centering} p{2cm}<{\centering} p{2cm}<{\centering} }
 \hline
 \multicolumn{4}{c}{Parameters of crystal field (eV)} \\
 \hline
 $2\mathcal{J}(3J_H)$ & $\Delta E_{e_g}$ & $10Dq$ & $\Delta E_{t_{2g}}$\\
 \hline
2.06 & 0.78 & 0.89 & 0.50 \\ 
 \hline
\end{tabular}

\begin{tabular}{ p{2.75cm}<{\centering} p{2.7cm}<{\centering} p{2.7cm}<{\centering}}
 \hline
 \multicolumn{3}{c}{Parameters of Energy Levels} \\
 \hline
 Orbital $\alpha$ & $E^{\uparrow}_{\alpha}$ (eV) & $E^{\downarrow}_{\alpha}$ (eV)\\
 \hline
$d_{3z^2-r^2}$ & 1.29 & 3.35 \\ 
$d_{x^2-y^2}$  & 0.51 & 2.57 \\ 
$d_{xz/yz}$    & 0.50 & 2.56 \\ 
$d_{xy}$       & 0.00 & 2.06 \\ 
 \hline
\end{tabular}
\caption{Energy range of characteristic peaks identified from the experimental RIXS spectrum within -0.5 eV to -4.5 eV. Fitting result of the crystal field parameters and the energy level on experimental data.}
\label{table:energy level}
\end{table}

In the second step, we use Eqs.~\eqref{eq:e1} - \eqref{eq:e4} to obtain the crystal field parameters $(\Delta E_{e_g}$, $\Delta E_{t_{2g}}$, 10$Dq)$, the intra-orbital SF energy 2$\mathcal{J}$, and the energy level parameter sets corresponding to each set of characteristic peaks. Note, in the above definitions $\Delta E_{e_g}$ and $\Delta E_{t_{2g}}$ represent the energy splitting of the $e_{g}$ and $t_{2g}$ orbitals. The energy splitting between the $e_{g}$ and the $t_{2g}$ orbitals is given by $10Dq$. The intra-orbital SF energy is given by $2\mathcal{J}$ where $2\mathcal{J} = E^{\downarrow}_{\alpha} - E^{\uparrow}_{\alpha}$. In the mean-field approximation the intra-orbital Hund’s exchange energy can be written as $\mathcal{H}^{MF}_{\mathrm{H}} = - J_{\mathrm{H}} \sum (1/2 + 2 S_{i\alpha}^z S_{i\beta}^z)$~\cite{Khomskii_2014}, where $J_H$ is the Hund's coupling. Based on this the Hund’s energy of the initial (final) state with four parallel spins (three parallel and one unparallel spins) is given by $E_i^{\mathrm{H}} = -6J_{\mathrm{H}}$ ($E_f^{\mathrm{H}} = -3J_{\mathrm{H}}$). Therefore, our definition of the intra-orbital SF energy can be related to the Hund's intra-orbital exchange energy as $2\mathcal{J} = 3J_{\mathrm{H}}$.
 
In the third step, we use the simulated energy levels as input to compute the RIXS intensity. We assume that when one of the electrons from the $e_{g}$ or the $t_{2g}$ orbital is excited, the other three electrons serve as spectators and the intermediate state has ultra-short survival time. Within this hypothesis, the cross sections from the ground state to all possible final states are calculated and the simulated spectrum as a function of energy loss $\hbar\omega$ is given by \begin{equation}
     I_{sim}(\hbar\omega)=\frac{1}{\pi}\sum_{\alpha,\beta}\sum_{\tau}\left[\frac{\Gamma_{\beta}W^{\pmb{\epsilon}_{out}}_{\pmb{\epsilon}_{in}}\left(d^{\uparrow}_{\alpha},d^{\tau}_{\beta}\right)}{(\hbar\omega+E^{\tau}_{\beta}-E^{\uparrow}_{\alpha})^2+\Gamma^{2}_{\beta}}\right ], 
\label{eq: Intensity}
\end{equation}
where we compute the spectra corresponding to each set of characteristic peaks. We insert the transition amplitudes $W^{\pmb{\epsilon}_{out}}_{\pmb{\epsilon}_{in}}\left(d^{\uparrow}_{\alpha},d^{\tau}_{\beta}\right)$ of the corresponding excitation channels. Next, we sum over all the spins ($\tau$) to consider both the NSF and SF channels. The orbital summation of the initial and final states ($\alpha$ and $\beta$) represents an exhaustive collection of all $dd$ excitation channels (that are symmetry allowed). Based on our calculation we find that the Lorentzian lifetime broadening $\Gamma_{\beta}$ is different between the NSF and the SF channels. We have $\Gamma_{NSF}=0.10$ eV and $\Gamma_{SF}=0.20$ eV. Furthermore, using the crystal field energy diagram shown in Fig.~\ref{fig:crystal feild}, we can deduce that $E^{\uparrow}_{\alpha} = E_{\alpha}-\mathcal{J}$ and $E^{\downarrow}_{\alpha} = E_{\alpha}+\mathcal{J}$. During the calculation process we considered a generalized parameter space in which anisotropy of the spin-flip interaction energy was taken into account. However, our calculations suggested that the energy $2\mathcal{J}$ arising from the transition between the orbitals (of the same type) do not differ much. They belong to the same energy range $\Delta E_4$ shown in Table~\ref{table:energy level}. Thus, we did not consider an anisotropic $2\mathcal{J}$ interaction in all our subsequent calculations.

Finally, the root-mean-squared-error (RMSE) between each set of simulated spectra (obtained by considering all possible combinations that can arise in the variation  of $\Delta E_i$ shown in Table~\ref{table:energy level}) and the experimental data was computed using \begin{equation}
    RMSE=\sqrt{\frac{1}{N}\sum_{i=1}^{N}\left[ I_{exp}\left(\hbar \omega_{i}\right)-I_{sim}\left(\hbar \omega _{i}\right)\right ]^{2}},  
    \label{equ:RMSE}
\end{equation} where $N$ is the total number of experimentally measured energy loss $(\omega_i)$ data points, $I_{exp}$ is the original experimental data, and $I_{sim}$ is the simulated spectrum. We report the simulated parameters  corresponding to the smallest RMSE values in Table~\ref{table:energy level}. In the next section, we discuss the experimental spectrum and its interpretation based on our simulated RIXS calculation. We also predict the RIXS spectra for the in- and out- of plane spin orientation and for various $\delta$ (relative to the z-axis) angles of the scattering wavevector $\bm{q}$. 

\section{SIMULATED RIXS spectrum}\label{sec:rixs spectrum}   

\begin{figure}[t]
\centerline{\includegraphics[width=7.5cm]{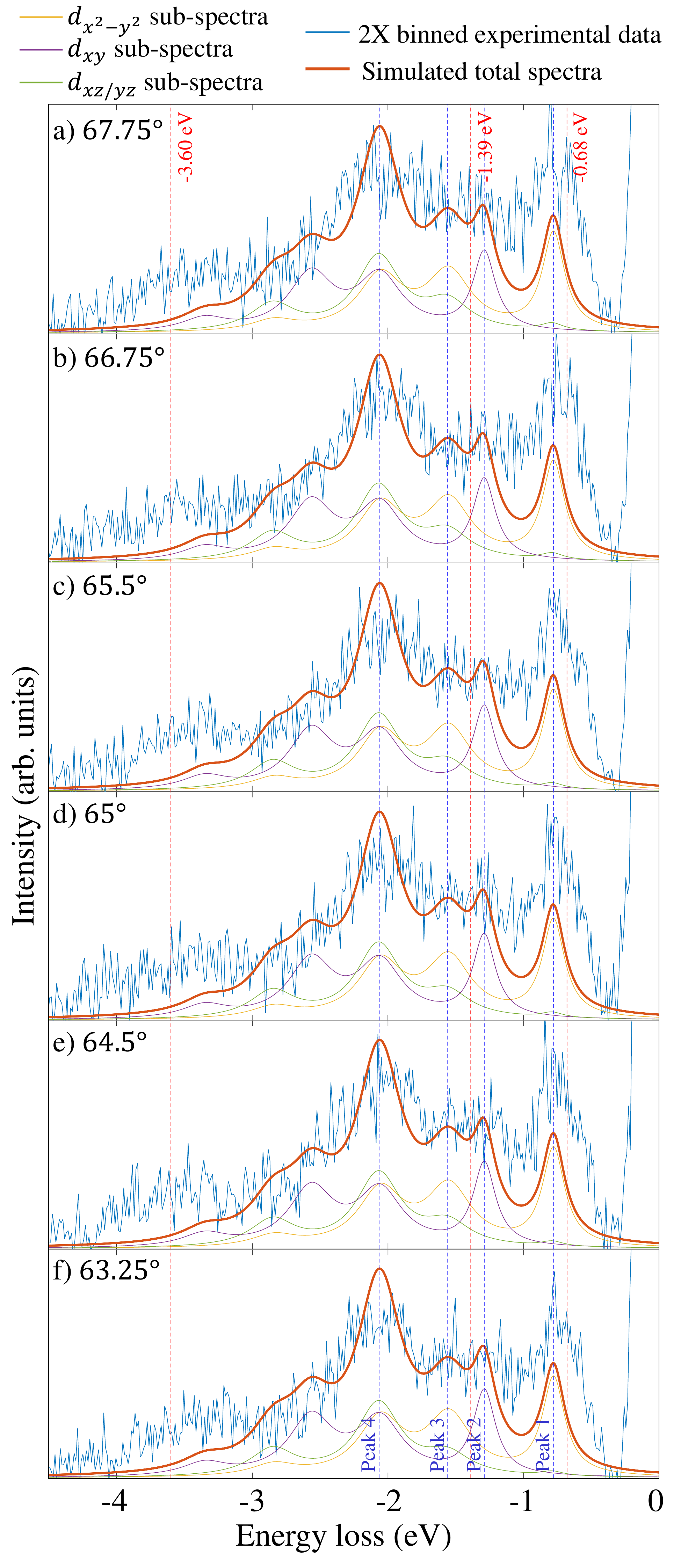}}
\caption{Simulated and experimental RIXS spectra of $dd$ excitations. The different panels refer to the RIXS intensity obtained by varying the scattering geometry angle $\theta$, see Fig.~\ref{fig: experiment from -12 to 2} for definition. The spin-flip interaction energy $2\mathcal{J}$ (which is equal to $3J_H$) is visible from the fourth peak $\Delta E_4$ of the simulated total spectra. We also show the location of peaks 1 through 4 for easy comparison (see text for discussion of these peaks).}
\label{fig: all rixs spectrum}
\end{figure}

\begin{figure}[t]
\centerline{\includegraphics[width=9.0cm]{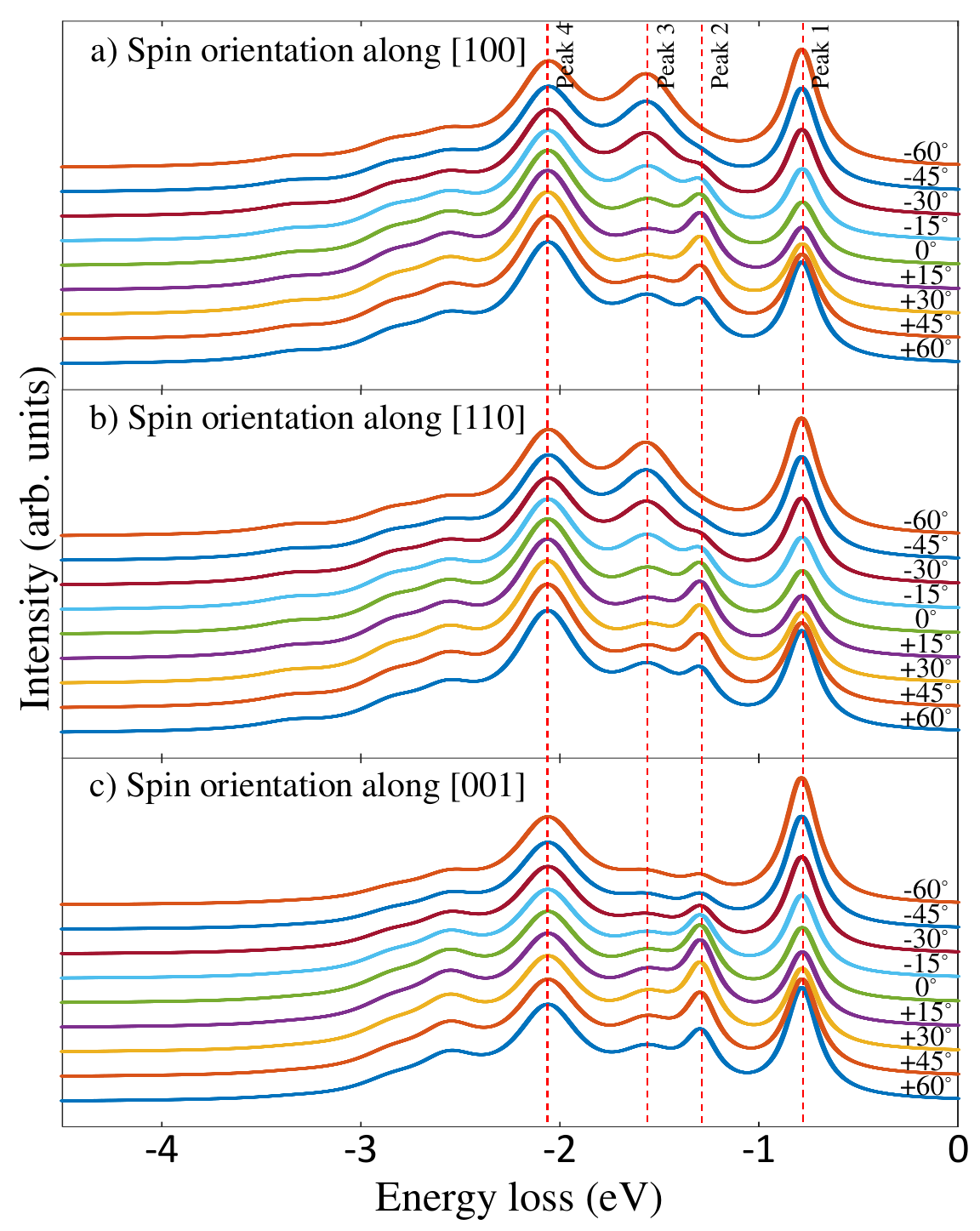}}
\caption{Theoretically predicted RIXS spectrum at Mn L$_{3}$ resonance in other experimental conditions (different spin orientation and angle $\delta$). The angle $\delta$ of each spectrum is listed on the right side of the figure. }
\label{fig: predicted spectrum}
\end{figure}
In Fig.~\ref{fig: all rixs spectrum}, we show the results of comparing the simulated spectra generated using the above fitting procedure with the experimental spectra. As evident from the plots, the physical mechanism underlying each peak on the sub-spectrum is a single $dd$ excitation channel. The sub-spectra of each different initial state are summed together to obtain the total RIXS spectra of all the channels. Comparing the theoretical calculations to the experimental data, we find that the trends in both the characteristic peaks and the scattering intensity changes coincide with each other. Therefore, our simulated spectrum is very close to the experimental spectrum. We were able to distinguish the energy boundaries of different excitations (NSF, SF, and charge-transfer) arising in the lightly manganese doped Sr$_{3}$Ru$_{2}$O$_{7}$ by identifying the sub-spectra (from our calculation in Sec.~\ref{sec:calculation} and Sec.~\ref{sec:parameter}) which contributes to the overall experimental RIXS intensity. These sub-spectra are denoted by thin solid lines in Fig.~\ref{fig: all rixs spectrum}. The vertical dashed lines demarcate the various excitation zones. The NSF $dd$ excitations occur from -1.39 to -0.68 eV. The SF $dd$ excitations takes place in the interval from -3.60 to -1.39 eV. The range of $dd$ excitation energy loss lies mainly in the interval from -3.60 eV to -0.68 eV. And it can be determined that the scattering intensity peak between -12.00 eV to -4.00 eV in Fig.~\ref{fig: experiment from -12 to 2} corresponds to the charge transfer excitation. However, in the low energy loss, the scattering intensities of the calculated results are generally lower than the experimental spectral lines. This is likely caused by the fact that the magnetic or phonon excitations are detected experimentally, whereas the calculated results only include the intensities of the $dd$ excitations.
Also, adjusting the scattering geometry parameters, we obtained the RIXS intensity at different angles $\theta$. Since the variation in $\theta$ is not substantial in the experimental setting the total RIXS spectra changes very slightly. 

In Fig.~\ref{fig: predicted spectrum}, we predict the RIXS spectra for $\delta$ variation between -60$^\circ$ to +60$^{\circ}$ for spin orientation that changes from in-plane ([100] and [110]) to out-of-plane ([001]), see Fig.~\ref{fig: scattering geometry}. We track the evolution of the four characteristic peaks to observe trends in the spectra. Irrespective of whether the spin orientation is in- or out-of-plane, we find that the intensity of the first peak gradually increases as $\delta$ changes from 0$^\circ$ to $\pm$60$^\circ$. The growth of the first peak arises from the $|d^{\uparrow}_{x^2-y^2}\rangle$ $\rightarrow$ $|d^{\uparrow}_{3z^2-r^2}\rangle$ transition, which gradually increases when the angle changes from 0$^\circ$ to $\pm$60$^\circ$, as seen in Fig.~\ref{fig:amplitudes}. When the spin is along the in-plane direction the spectral weight shifts from the third to the second peak as $\delta$ varies. The second (third) peak is higher in intensity than the third (second) peak from 0$^\circ$ to 60$^\circ~(-60^\circ)$. At $\delta$ = -60$^\circ$, the second peak cannot be distinguished in the total RIXS spectrum. When the spin is oriented out-of-plane, there is a gradual increase in the intensity of both the second and the third peaks over the entire range of $\delta$ variation. However, the second peak has a greater change. For a similar value of $\delta$, the intensity of the fourth peak is consistently greater for the in-plane orientation compared to the out-of-plane spin direction. This behavior can be tracked to the presence of more intra-orbital SF excitation channels when the spin is in-plane compared to its out-of-plane configuration. As shown in Fig.~\ref{fig:amplitudes}(c), the inter-orbital SF channels from $d_{xy}$ and $d_{x^2-y^2}$ orbital are suppressed when the spin orientation is along [001].

\section{Conclusions}\label{sec:conclusion} 
Utilizing experimental data and a single-ion based RIXS model, we simulate the $dd$ excitation spectra of lightly manganese doped Sr$_{3}$Ru$_{2}$O$_{7}$.  We have determined the $dd$ excitation energies of the Mn$^{3+}$ ion at the $L_3$-edge which are directly related to the energy level structure of the $3d^4$ in the $D_{4h}$ crystal field symmetry. Based on our calculation, we have identified the NSF, SF, charge-transfer boundaries in the experimental RIXS spectrum. Using a self-consistent fitting procedure we have obtained the crystal field parameters, the single spin-flip interaction energy, and the energy levels of the $d$ orbitals of the Mn$^{3+}$ ion embedded in the matrix of the Sr$_{3}$Ru$_{2}$O$_{7}$ compound. The results of our fit are reported in Table~\ref{table:energy level}. We can calculate the superexchange constant using the free ion Racah parameters~\cite{PhysRevB.51.13942, doi:10.1143/JPSJ.9.766}. Using the expression $J=4B +C $ ~\cite{maekawa2013physics}, we find $2\mathcal{J}=2.06$ eV, which is exactly the value we found from the experimental data. This confirmation lends support to the hypothesis mentioned in the introduction that the diluted Mn$^{3+}$ ions in the host compound can be treated as a free ion. Additionally, it bolsters confidence in our fitting procedure and in the prediction of the other material parameters of free Mn ion, including the excitation energy boundaries.  

Utilizing the material parameters from the fit, we can predict the RIXS spectra for other scattering geometry set-up that have not been explored in the current experimental setting. We predict the RIXS spectra for spin orientation that changes from the in-plane to the out-of-plane configuration. We find that as the spin direction evolves there is a significant decrease in the number of the intra-orbital SF channels. We find some non-trivial intra-orbital SF excitation channels for both non-cross polarization $(\pmb{\sigma}, \pmb{\sigma})$. Apparently, these quantum transitions defy the conventional logic of angular momentum conservation. A similar effect was predicted earlier to occur within the context of magnons found in the copper based Kagome compound~\cite{xiong2020resonant}. In our experiment, as the spin orientation changes from in-plane to out-of-plane, these channels appear in different orbital transitions and distinct polarization channels. In the in-plane configuration, the non-trivial intra-orbital SF originates from the $d_{xy}$ orbital in the $\pmb{\sigma}_{in}-\pmb{\sigma}_{out}$ channel. But, when the spin is out-of-plane, the non-trivial intra-orbital SF happens from the $d_{xz/yz}$ orbital in the $\pmb{\pi}_{in}-\pmb{\pi}_{out}$ channel. The preference for out-of-plane spin orientation to choose the $\pmb{\pi}_{in}-\pmb{\pi}_{out}$ channel was also observed on the $d_{3z^{2}-r^{2}}$ orbital in the non-collinear non-coplanar Kagome copper compound~\cite{xiong2020resonant}. Finally, we hope that our work will stimulate the RIXS community to start exploring manganite RIXS (and more complex ionic configurations) which promise to host both exciting fundamental physics (conservation of angular momentum) and an opportunity to determine materials parameters. 

\begin{acknowledgements}
This project is supported by NKRDPC-2022YFA1402802, NKRDPC-2018YFA0306001, NSFC-92165204, NSFC-11974432, Shenzhen International Quantum Academy (Grant No. SIQA202102), and Leading Talent Program of Guangdong Special Projects (No. 201626003). The experiments have been performed at the ADRESS beamline of the Swiss Light Source at the Paul Scherrer Institut (PSI). The work at PSI is supported by the Swiss National Science Foundation through project no. 200021\_178867, and the Sinergia network Mott Physics Beyond the Heisenberg Model (MPBH) (SNSF Research Grants CRSII2\_160765 and CRSII2\_141962). T.C.A. acknowledges funding from the European Union’s Horizon 2020 research and innovation program under the Marie Skłodowska-Curie grant agreement No. 701647 (PSI-FELLOW-II-3i program). Single crystals were grown and  provided by Yoshiyuki Yoshida of AIST (Japan). YDC is supported by Advanced Light Source, a DOE Office of Science User Facility under contract no. DE-AC02-05CH11231. B.F. acknowledges Welch Professorship support from the Welch Foundation (Grant No. L-E-001-19921203), University of Houston and the Texas Center for Superconductivity. T. D. acknowledges the hospitality of KITP at UC-Santa Barbara. A part of this research was completed at KITP and was supported in part by the National Science Foundation under Grant No. NSF PHY-1748958. TD acknowledges useful discussions with Theja DeSilva.
\end{acknowledgements}

\bibliography{Ref}
\end{document}